\documentclass[prl,aps,preprintnumbers,amsmath,amsfonts,twocolumn]{revtex4}
\usepackage{graphics,bm}
\usepackage{epsfig,graphicx}
\usepackage[english]{babel}
\usepackage{amsfonts}
\usepackage{amsmath}
\usepackage{latexsym}
\usepackage{epsfig}
\usepackage{graphics}
\usepackage{graphics,bm}
\usepackage{subfigure}
\usepackage{dcolumn}
\usepackage{bm}
\usepackage{rotating}
\usepackage[usenames]{color}

\def\br{{\bf r}}

\def\bk{{\bf k}}

\def\Rk{R( {\bf k} )}
\def\Rkz{R_0} 

\def\velv{{\bf v}_{ s}}
\def\veln{{\bf v}_{ n}}

\def\vk{v_{\bf k}\left(\br\right)}
\def\uk{u_{\bf k}\left(\br\right)}
\def\zk{z_{\bf k}\left(\br\right)}
\def\vkstar{v_{\bf k}^*\left(\br\right)}
\def\ukstar{u_{\bf k}^*\left(\br\right)}
\def\vkquad{v_{\bf k}^2\left(\br\right)}
\def\ukquad{u_{\bf k}^2\left(\br\right)}

\def\zkstarmen{z_{{\bf k}}^*\left(\br\right)}

\def\ndi{ n_{\rm R}}
\def\ndis{ n_{\rm R}\left(\br\right)}
\def\ndisz{ n_{{\rm R}_0}\left(\br\right)}
\def\ndisZ{ n_{{\rm R}_0}}

\def\ndisoz{ n_{{\rm R}_0}\left(\bf 0\right)}
\def\Ndis{N_{\rm R}}
\def\Ndisz{N_{{\rm R}_0}}
\def\nr{n\left(\br\right)}
\def\nbog{n_{\rm Bog}\left(\br\right)}

\def\nTF{n_{\rm TF}\left(\br\right)}
\def\ntf{n_{\rm TF}}
\def\nz{n_0\left(\br\right)}
\def\nzz{n\left(\bf 0\right)}

\def\deltan{\delta n}
\def\deltanstar{{\delta n}^{\ast}}
\def\sol{\delta n\left(\br\right)}

\def\mur{ \mu_{{\rm R},l}\left(n\right)}

\def\Ek{\hbar\Omega_{{\bf k}}\left(\br\right)}

\def\ddt{\frac{\partial}{\partial t}}

\def\wo{\omega_{\rm HO}}
\def\aosc{a_{{\rm HO}}}
\def\mul{\mu_{l}}
\def\Vext{V_{\rm ext}\left(\br\right)}
\def\Vexti{V_{\rm ext}}
\def\tdms{\int d \br}
\def\tdmss{\int\!\! d \br}

\def\Rtfu{R_{\rm TF}}
\def\mTFz{\mu_{\rm TF}}
\def\mTFu{\mu_0}
\def\effemu{{f}_{\mu}}
\def\effeR{{f}_{\rm R}}
\def\effeX{{f}_{\chi}}
\def\xiti{\tilde{\xi}}
\def\riti{\tilde{r}}
\def\omegaperp{\omega_{\bot}}
\def\rperp{r_{\bot}}

\begin{document}
\flushbottom
\raggedbottom

\title{
Collective Oscillations in Trapped Bose-Einstein Condensed
Gases in the Presence of Weak Disorder}

\author{G.M. Falco, A. Pelster, R. Graham}

\affiliation{Universit\"at Duisburg-Essen, Fachbereich Physik, Campus Duisburg
\\Lotharstrasse 1, 47057 Duisburg, Germany }

\begin{abstract}
The influence of
a weak random potential on the collective modes 
of a trapped interacting
Bose-Einstein condensate at zero temperature is calculated 
in the limit when the correlation length of the disorder is smaller than the healing length
of the superfluid. 
The problem is solved in the Thomas-Fermi limit
by generalizing 
the 
superfluid hydrodynamic equations
to allow for the presence of weak disorder.
We find that the
disorder-induced frequency-shifts of the low-energy excitations
can be  
of the same order of magnitude as the beyond mean-field corrections in the normal interaction
recently observed experimentally.

\end{abstract}
\maketitle

\section{Introduction}  
Some time ago, Huang and Meng \cite{Huang92} have studied
a homogeneous three-dimensional hard-sphere Bose gas in a random external potential
as a model of superfluid helium in a disordered medium \cite{Reppy90}. 
In qualitative agreement with the experiments in porous media,
they found that 
the formation of 
local condensates in the minima of the random potential  reduces the superfluid component
of the fluid even at zero temperature, where, in the absence of disorder,
the whole fluid 
would be superfluid \cite{Kalatnikov62}. 
The 
recent experimental advances
in trapping Bose-Einstein condensates
in a disordered medium \cite{Lye05,Clement05}
makes it interesting to test in a more quantitative manner 
the predictions of the model considered by Huang and Meng.
For this purpose, we extend the latter approach to 
include a harmonic trapping potential in addition to the 
weak external random potential.
In the present work we consider a condensate in 
the limit of a large number of particles $N$ and in the presence of
disorder with a correlation length
shorter than the healing length of the superfluid.
These conditions allow for 
a simple hydrodynamical formulation of the problem similar to the 
theory of wave propagation in random elastic media \cite{Ishimann78}. 

In the case of weak disorder,
the corrections 
in the density profile or in the released energy 
of the Huang-Meng theory 
\cite{Huang92} turn out to be 
too small to be measurable.
Nevertheless, 
in the present paper we argue that 
disorder-induced shifts might be observable in the frequencies of the collective excitations
\cite{Stringari96},
because these can be measured with 
an accuracy of a few fractions of a percent \cite{Kurn98}.
More precisely,
we show that long wavelength  
disorder can shift the frequencies of the low-energy excitations
with the same order of magnitude but opposite sign as the beyond 
mean-field shifts 
due to 
repulsive atomic interactions \cite{Pitaevskii98} in the gas.
The latter effect has been recently observed experimentally by the Innsbruck group \cite{Grimm06}. 
Furthermore,
in a harmonic trap, the influences of disorder and interaction should be readily distinguishable.
This follows because, according to the generalized 
Kohn's theorem \cite{Dobson94,Kohn61},
the interaction cannot alter the frequency of the lowest dipole mode \cite{Pitaevskii98} while,
as we show, the
latter is shifted by the presence of a weak random external potential.

\section{Hydrodynamic equations} 
We consider 
a harmonically trapped Bose gas in
an external random field.
The grand-canonical Hamiltonian of the system is
\begin{equation}
\label{eq:actionmomspace}
K\!=\!\tdmss
\left\{\psi^{\dagger}{\Big{[}}-\frac{\hbar^2\nabla^2}{2m}-\mu+U+\Vexti{\Big{]}}\psi
+\frac{g}{2}
{\psi^{\dagger}}^2\psi^{2}\right\}
\end{equation}
where $\psi^{\dagger}(\br)$, $\psi(\br)$ are the field operators of an atom with mass $m$, $\mu$
is the chemical potential, 
and the interaction at low energy is described 
by the two-body ${\rm T}-$matrix
$g=4\pi\hbar^2 a/m$ in terms of the $s-$wave scattering length $a$.
The harmonic trapping potential is
$\Vext=m\left(\omega_x^2 x^2+\omega_y^2 y^2+\omega_z^2 z^2
\right)/2$. The oscillator frequencies define the harmonic oscillator length
$\aosc=\left(\hbar/m\wo\right)^{1/2}$, where 
$\wo=\left(\omega_x\omega_y\omega_z\right)^{1/3}$.
The disorder potential $U$ is chosen 
with a Gaussian probability distribution
characterized by the ensemble averages
$\langle U({\bf r})\rangle=0$ and 
$\langle U({\bf r})U({\bf r}')\rangle=R\left({\bf r}-{\bf r}'\right)$,
with $R\left({\bf r}\right)=\int d{\bk}/(2\pi)^3 e^{i\bk \br}\Rk$.
In the case of
a fast-decaying disorder-correlation $R(\bf r)$, the results of the theory do not depend 
significantly on its shape \cite{Timmer06}. 
We consider the 
case of a Gaussian correlation
$\Rk = R_{0} \,e^{- {k}^2 \xi^2/2}$,
where $R_{0}$ and $\xi$ characterize
the strength
and the correlation length of the disorder, respectively.

A random potential in a Bose-Einstein condensate 
causes incoherent scattering which tends to localize and to deplete the 
condensate wavefunction. 
For a statistically homogeneous system, Huang and Meng have shown \cite{Huang92} that,
even at $T=0$, 
this results also in a depletion
of the {\it superfluid} density $n_s$.
This latter must be distinguished from the {\it condensate} density $n_0$
and is related to the total density $n$ by the relation $n_s=n-n_n$,
where the normal (i.e. non-superfluid) component of the gas $n_n$
can be related to the disorder-induced depletion $\ndi$
of the condensate through
$n_n=({4}/{3})\ndi$ \cite{Huang92}.
Consistently with this picture, we assume that for weak disorder and at $T=0$
the superfluid component of the gas
can be described by the phenomenological two-fluid hydrodynamic equations \cite{Kalatnikov62}
\begin{align}
\label{eq:hydrodeqts}
&\frac{\partial}{\partial t}n+{\bf{\nabla}}\left(\velv n_s+\veln n_n\right)=0\nonumber\\
&m \ddt \velv+\nabla\left(\mu+\frac{1}{2}m\velv^2\right)=0.
\end{align}
The hydrodynamical variables in this description are assumed to be ensemble averages 
over the realizations of the disorder potential.
Therefore, the physical validity of Eqs.~(\ref{eq:hydrodeqts}) must be considered as being 
restricted to the self-averaging regime
where the wavelength $q^{-1}$ of the hydrodynamic modes
is much larger than the length-scale $\xi$ of
the disorder potential. 
This condition determines the range of validity of our theory.
The equations~(\ref{eq:hydrodeqts}) can be understood as a Landau ``two-fluid" model,
for the unpinned part of the condensate as superfluid component and the localized pinned 
condensate as normal component.
We use these equations to describe the collective excitations
of energy $\hbar\omega\approx \hbar\wo\ll\mu$ in the presence of the harmonic trap.
In the Thomas-Fermi regime and for long-wavelength oscillations \cite{Dalfovo99}, the
non-uniform system can be considered locally as homogeneous,
with the space-time-dependent density $n\left(\br,t\right)$ and the chemical potential
related by
$\mu(\br,t)=\mul\left[n\left(\br,t\right)\right]+\Vext$
in the local density approximation \cite{Dalfovo99}.
Here $\mul\left[n\left(\br,t\right)\right]$
is the chemical potential of a uniform gas at density $n\left(\br,t\right)$.
Because of the pinned character of the normal component, we must furthermore assume that 
only
the superfluid component reacts to the probe, while the pinned normal
component remains stationary. 
This situation is reminiscent of
the physically closely related problem of the fourth sound in $^4$He \cite{Kalatnikov62}
and is expressed by the condition $\veln=0$ in Eqs.~(\ref{eq:hydrodeqts}). 

Decomposing 
$n\left(\br, t\right)=n\left(\br\right)+\delta n\left(\br,t\right)$ with the time-independent background density
$n\left(\br\right)=n_s\left(\br\right)+n_n\left(\br\right)$,
and $\mu\left(\br,t\right)=\mu_0+\delta\mu\left(\br,t\right)$ with $\delta
\mu=\left(\partial\mu/\partial n\right)\delta n$, and restricting ourselves to the linear regime,
Eqs.~(\ref{eq:hydrodeqts}) lead to the wave-equation
\begin{align}
\label{eq:hydrodeqtsbis}
&m \frac{\partial^2}{\partial t^2}\delta n\left(\br, t\right)-{\bf{\nabla}}
\left[n_s\left(\br\right)\nabla\left(\frac{\partial\mul
\left[n\left(\br\right)\right]}{\partial n\left(\br\right)}
\delta n\left(\br, t\right)\right)\right]=0.
\end{align}
The disorder and the trap both appear implicitely in Eq.~(\ref{eq:hydrodeqtsbis}) via the superfluid density $n_s(\bf r)$ 
and the equation of state 
$\mul\left[n\left(\br\right)\right]$ which are determined in the following two sections, respectively.
In the limit of a statistically uniform gas, 
Eq.~(\ref{eq:hydrodeqtsbis}) reduces to the wave-equation in 
a medium with a random refractive index \cite{John83}.

\section{Mean-field  theory in the trap} 

In order to complete the equation of motion~(\ref{eq:hydrodeqtsbis})  we have to determine both
the equilibrium superfluid density and the equation of state.
To this purpose we now  extend the Huang-Meng theory to a non-uniform system, by making use of
the local density approximation.
Expanding locally in plane waves $\Psi(\br)=(1/\sqrt{V})\sum_{\bk}a_{\bk}(\br) e^{i\bk\br}$,
the presence of the condensate is taken into account by setting 
$a_{{\bf k}}({\bf r})=\langle \sqrt{n_{{\bf 0}}({\bf r})V} \rangle\delta_{{\bf 0},{\bf k}}+\delta a_{{\bf k}}({\bf r})$. 
Retaining only the quadratic terms in the excitations $\delta a_{{\bf k}}$, 
$\delta {a^{\dagger}}_{{\bf k}}$ from the condensate, the truncated Hamiltonian
obtained from Eq.~(\ref{eq:actionmomspace})
can be diagonalized  by the Bogoliubov transformation \cite{Huang92}
$\delta a_{{\bf k}}({\bf r})=\uk\alpha_{{\bf k}}-\vk
\alpha_{{-\bf k}}^{\dagger}-\zk$ and 
$\delta a_{{\bf k}}^{\dagger}({\bf r})=\ukstar\alpha_{{\bf k}}^{\dagger}-\vkstar
\alpha_{{-\bf k}}-\zkstarmen$,
where the coherence factors $u_{{\bf k}}$ and $v_{{\bf k}}$ 
and the complex numbers $z_{{\bf k}}$ can be taken real by appropriately choosing  
the phase of the complex fields.
Then we have 
$\ukquad=\{1+{\left[\epsilon_{\bf k}-\mu
+\Vext+2g\nr
\right]}/{\Ek}\}/2$,
$\vkquad=
\ukquad-1$, and
$\zk=\left[
\nr V\right]^{\frac{1}{2}}
{U_{\bf k}}
\left[
\uk-\vk\right]^2/\Ek$,
with $\epsilon_{\bk}=\hbar^2 k^2/2m$
and, where
the Bogoliubov spectrum \cite{Bogoliubov47} is given by 
$\Ek=
\sqrt{\left[\epsilon_{\bf k}-\mu+\Vext+2g
\nr \right]^2-\left[g
\nr\right]^2}$.
The density is given by
$\nr=\nz+\nbog+\ndis$,
where, besides the condensate density, $\nz$ the local Bogoliubov depletion density \cite{Bogoliubov47} 
$\nbog=(8/3)
\nr\sqrt{
\nr a^3/\pi}$ appears, 
and the depletion $\ndis$ induced by the random potential  
\begin{align}
\label{eq:disdepldelt}
\ndis=\ndisz\effeR[4\pi
\nr \xi^2 a].
\end{align}
Here,
$\ndisz=\Rkz \left(m^2/8 ~\pi^{3/2}~
\hbar^4\right)\sqrt{
\nr/a}$
denotes the local depletion in the limit of a $\delta$-correlated disorder potential
and the function \cite{Kobayashi02}
\begin{align}
\label{eq:effeRR}
\effeR(x)=\left[
e^{2 x}\left(1+4 x\right) {\rm Erfc}\left(\sqrt{2 x}\right)-2\sqrt{{2 x}/{\pi}}
\right]
\end{align}
(with $\effeR(0)=1$) 
includes the effects of its non-vanishing correlation length $\xi$.
With $\ndis$ now in hand, the local superfluid density $n_s({\bf r})$ is given 
in terms of $n({\bf r})$ by the relations
$n_s({\bf r})=n({\bf r})-n_n({\bf r})$,  where the normal component 
is $n_n({\bf r})=(4/3)\ndis$ as discussed before.

The total disorder-induced depletion of the condensate $\Ndis=\tdms \ndis$,  
can be calculated to leading order in $\Rkz$ 
by replacing in Eqs.~(\ref{eq:disdepldelt}) 
the density $n({\br})$ by its zero-order 
Thomas-Fermi approximation 
$\nr\simeq\left[\mTFz-\Vext\right]/g$, where $\mTFz=\frac{\hbar \wo}{2}
(\frac{15 N a}{\aosc})^{2/5}$ is the mean-field chemical potential
of the Gross-Pitaevskii theory. 
In the case of $\delta$-correlated disorder,
integrating over the mean-field
Thomas-Fermi radius 
we get
${\Ndisz}/{N}\simeq({15\pi}/{32})[{\ndisoz}/{n\left(\bf 0\right)}]$.
The theory is valid when $\Ndisz/N\ll 1$, which can also be rewritten as 
$\Rkz'\left(\br\right)
\equiv m^2 \Rkz/8 \pi^{3/2}\hbar^4 
\sqrt{n\left(\br\right) a}\ll 1$, or, in local form as
 the condition $\ndisz \ll n\left(\br\right)$. 
In the Thomas-Fermi regime the latter condition is satisfied everywhere in the trap with the exception
of a narrow shell 
at the boundaries 
where the condensate density vanishes. 
In the present work we neglect possible small corrections from this effect. 

\section{Beyond mean-field equation of state} 

The ground-state density $n\left(\br\right)$
at equilibrium in Eq.~(\ref{eq:hydrodeqtsbis}) 
can be calculated by using the local density approximation for the chemical potential 
$\mu_0=\mul\left[n\left(\br\right)\right]+\Vext$.
In the mean-field Huang-Meng theory outlined in the previous section,
the latter is determined
by the Thomas-Fermi result $\mTFz=g n({\bf r})+\Vext$
of the Gross-Pitaevskii theory.
In this mean-field approximation 
the effects
of quasi-particle interactions \cite{Huang57} as well as the scattering
between the quasi-particles and the impurities \cite{Lopatin02} are neglected. 
In order to incorporate these processes in the calculation of the collective modes,
the ground-state value of the chemical potential $\mu$  
has to be fixed 
in terms of $N=\tdms n({\bf r})$
by including beyond mean-field corrections in the equation of state.
These latter can be calculated in the framework of the Bogoliubov theory \cite{Falco07}
resulting in 
\begin{align}
\label{eq:muLopTzer}
\mul\left(n\right)
=& n g\left(1+ \frac{32}{3}\sqrt{\frac{n {a}^3}{\pi}}\right)+\mur,
\end{align}
where
$\mur=6g
\ndisz 
\effemu[4\pi
\nr\xi^2 a]$,
and  
$\effemu(x)=\left[
e^{2 x}\left(3+4 x\right) {\rm Erfc}\left(\sqrt{2 x}\right)-2\sqrt{{2 x}/{\pi}}
\right]/3$.
In the following we wish to focus on the effects of disorder. Therefore,
in the equation of state~(\ref{eq:muLopTzer}) 
we can neglect the beyond mean-field corrections 
due to the normal interactions,
keeping only those due to the disorder.
This is possible because, although the two different corrections can be of the same order, 
they lead to effects which are additive to lowest order.  
Moreover, 
in an experiment one could focus on the effects of the disorder by 
tuning, via a  Feshbach resonance,
to the regime $\Rkz'({\bf 0})\gg \sqrt{n({\bf 0})a^3}$ 
where they become dominant.
Other corrections due to finite-size, non-linearity and temperature have been
sufficiently discussed in Ref. \cite{Pitaevskii98} in connection with the
frequency shifts
induced by the beyond mean-field effects in the theory of Huang and Yang \cite{Huang57}.

Using Eq.~(\ref{eq:muLopTzer}), for small disorder, we can find the equation
for the ground-state density by iteration.
Defining the Thomas-Fermi density as 
$\nTF=\left[\mTFu-\Vext\right]/g$, 
with $\mTFu(N)$  
determined from the normalization condition $N=\tdms n({\bf r})$
{\it including} the 
disorder correction in the equation of state~(\ref{eq:muLopTzer}),
we find
\begin{align}
\label{eq:TFdensit}
\nr\simeq\nTF-6~\ndisz\effemu\left[4\pi\nTF \xi^2 a\right].
\end{align}
In the derivation we have simplified the disorder-induced depletion
$\ndisz$ of Eq.~(\ref{eq:disdepldelt})
via replacing $n$ by its zero-order approximation $\ntf$.
The same simplification is made
in the argument of $\effeR$ and will be used in that of the function $\effeX$ defined below. 

Another quantity we need to evaluate in the hydrodynamic equation~(\ref{eq:hydrodeqtsbis}) 
is the term proportional to the inverse of the compressibility.
From Eqs.~(\ref{eq:disdepldelt}) and~(\ref{eq:muLopTzer})
we have 
\begin{align}
\label{eq:dmuxifin}
\left(\partial\mul/\partial n\right)
=&g\{
1+
3[{\ndisz}/{\nTF}]\effemu\left[4\pi\nTF \xi^2 a\right]
\nonumber\\
&
+6[\ndisz/\nTF] 
\effeX\left[4\pi\nTF \xi^2 a\right]\},
\end{align}
with 
$\effeX(x)=x\left[\frac{2}{3}
e^{2 x}\left(5+4 x\right) {\rm Erfc}\left(\sqrt{2 x}\right)-\frac{4}{3}\sqrt{\frac{2}{\pi}}
\frac{1+x}{\sqrt{x}}
\right]$.

\section{Collective modes}

Having calculated the beyond mean-field corrections due to disorder in the equation of state,
we can proceed to determine the explicit form of the hydrodynamic equation
describing the low-energy collective modes of the system in the linear regime. 
Using the result of Eq.~(\ref{eq:TFdensit}) and Eq.~(\ref{eq:dmuxifin}), and retaining
only terms linear in $R_0$, the hydrodynamic equation of Eq.~(\ref{eq:hydrodeqtsbis})
can be put into the final form
\begin{align}
\label{eq:hydrodeqtrisxi}
m \frac{\partial^2}{\partial t^2}\deltan-&{\bf{\nabla}}
\left[g\ntf\nabla\deltan
\right]=
-{\bf{\nabla}}\left[\frac{4 \ndisZ
\effeR
}{3}g{\bf{\nabla}}
\deltan
\right]
\\
&-\nabla^2\left\{g\left[3~\ndisZ\effemu
-6\ndisZ
\effeX
\right]
\deltan
\right\},
\nonumber
\end{align}
where we have left the $\br-$dependence of the coefficients implicit.
In the limit of a uniform gas 
the solutions of Eq.~(\ref{eq:hydrodeqtrisxi}) 
exhibit a phonon dispersion $\hbar\omega=c q$. 
Using Eq.~(\ref{eq:TFdensit}) and expanding for $\xi\ll \xi_{\rm heal}$ we find  
$c^2\simeq c_0^2
\left[1+(\frac{5}{3}-
\frac{32}{3\sqrt{\pi}}\frac{\xi}{\xi_{\rm heal}}
)\frac{ n_{\rm R}}{ n}\right]$
which describes the shift of the velocity of sound induced by  
the disorder from the Bogoliubov
result $c_0^2=g n/m$ 
of the clean system with healing length $\xi_{\rm heal}=1/\sqrt{8\pi n a}$. When $\xi=0$  
this reproduces the result found in Refs. \cite{Lopatin02,Giorgini93}
for $\delta-$correlated disorder.
Moreover, putting the r.h.s. of Eq.~(\ref{eq:hydrodeqtrisxi}) equal to zero we recognize the mean-field
equation used by Stringari in Ref. \cite{Stringari96}
in the case of an isotropic trap.
This yields
the dispersion relation
$\omega_0\left(n_r,l\right)=\wo\left(2n_r^2+2n_r l+3 n_r+l
\right)^{1/2}$, 
for excitations with $n_r$ radial nodes and $l$ angular momentum.
In the presence of disorder, Eq.~(\ref{eq:hydrodeqtrisxi}) can be solved considering the right-hand side as a
small perturbation. 
Defining the function $h(\br)=3~\ndisz\effemu\left[4\pi\xi^2 \nTF a\right]
-6\ndisz\effeX\left[4\pi\xi^2 \nTF a\right]$,
we find ultimately the frequency shift
\begin{align}
\label{eq:finaleq}
\frac{\delta\omega_0 (\xi)}{\omega_0}\!\!=\!\!
\frac{g}{2 m \omega_0^2 }
\frac{\tdms [\left(\nabla^2\deltanstar\right)  h
\deltan
+\deltanstar\nabla\left(\frac{4}{3}\ndisZ
\effeR
\nabla\deltan
\right)]}{\tdms\,\,\deltanstar\deltan},
\end{align}
where $\delta n$ are the solutions of the mean-field equation. 
In contrast with the beyond mean-field effects due to the atomic interaction
described in Ref.~\cite{Pitaevskii98},
the frequencies of the so-called surface-modes are influenced by disorder. 
These modes have the principal quantum number $n_r=0$
and represent solutions of the type
$\sol\sim r^l Y_{lm}$ with a mean-field dispersion law given by 
\cite{Stringari96}
${\omega_0}(n_r=0)
=\wo \sqrt{l}$.
Because they satisfy the condition $\nabla^2\sol=0$
the left term in the numerator of the r.h.s. of Eq.~(\ref{eq:finaleq}) is zero.
Note that this term comes from the change in the macroscopic compressibility
contained in the term $\partial\mu/\partial n$ in the equation of state of 
Eq.~(\ref{eq:hydrodeqtsbis}).
However, the right term carrying the factor $4/3$ does not
vanish. This correction originates because
the normal part of the liquid remains stationary under an external probe.
In general 
we obtain for the frequency shift of the surface modes
\begin{align}
\label{eq:gensurffinxix}
\frac{\delta\omega_{0}(\xi)}{\omega_0}=
\frac{2l+3}{3} \Rkz'\left(\bf{0}\right) I_{0,l,m}\left(\xiti\right)
\end{align}
with
\begin{align}
I_{0,l,m}\left(\xiti\right)
=2\int_{0}^1 d\riti \riti^{2l+1}\frac{\partial}{\partial \riti}
\left[
\left(1-\riti^2\right)^{1/2}\effeR\left(\xiti^2(1-\riti^2)\right)\right]\nonumber
\end{align}
independent from the quantum number $m$.
The dimensionless variable $\xiti$ is given by the relation 
$\xiti^2=\left({\xi\Rtfu}/{\aosc^2}\right)^2/2$.
In the case of the dipole mode $\deltan\sim r \cos\theta$
with $l=1$, $m=0$,
the mean-field result $\omega_{\rm dip}=\wo$ coincides
with the harmonic oscillator result $\omega_{\rm osc}=\wo\left(2n_r+l\right) $. This follows
from the fact that in a harmonic potential the lowest dipole mode 
($n_{\rm r}=0$, $l=1$)
corresponds to the oscillation 
of the center of mass, and is unaffected by the interatomic forces (Kohn's theorem) \cite{Kohn61}.
However, during its motion, the superfluid density is ``hampered" by the normal 
component which remains stationary.
In the limit of $\delta-$correlated disorder,
Eq.~(\ref{eq:gensurffinxix}) gives 
${\delta\omega_{\rm dip}(\xi=0)}/{\omega_{\rm dip}}
=-({5\pi}/{16})\Rkz'\left(\bf{0}\right)$. In the same limit, the shift of 
the quadrupole mode $l=2$, $m=2$, described by $\deltan\sim r \sin^2{\theta}e^{2i\phi}$,
is
${\delta\omega_{\rm Q}(\xi=0)}/{\omega_{\rm Q}}=-({35\pi}/{96})\Rkz'\left(\bf{0}\right)$.
According to the definition of the dimensionless parameter $\Rkz'\left(\bf{0}\right)$, we see that, besides disorder, also the presence
of interactions and thus of superfluidity, is crucial in order to have the effect. 

\begin{figure}
\begin{center}
\vskip 1 cm
\resizebox{7cm}{4.6cm}{
 \includegraphics{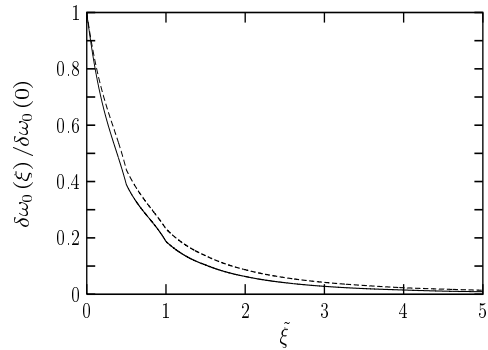}
 } 
\end{center}
  \caption{ Relative frequency shift $\delta\omega_{0}(\xi)/\delta\omega_{ 0}(0)$
  as a function of the variable $\xiti={\xi\Rtfu}/{\aosc^2}
\sqrt{2}$ 
  in the case of the surface excitations with $l=1$ (solid line) and $l=2$ (dashed line).
} 
  \label{figure1}
\end{figure}

The effects of a non-zero disorder correlation length $\xi$ 
on the frequency shift of the surface modes
can be better understood
when considering the ratio 
\begin{align}
\label{eq:gensurffinxi}
\frac{\delta\omega_{0}(\xi)}{\delta\omega_{ 0}(0)}=-
\frac{2}{\sqrt{\pi}}\frac{\Gamma(l+2)}{\Gamma{(l+3/2)}}
I_{0,l,m}\left(\xiti\right)
\end{align}
obtained from Eq.~(\ref{eq:gensurffinxix}).
In Fig. \ref{figure1} the result is illustrated in the case of the dipole and the quadrupole oscillations.
We plot the relative frequency shift
$\delta\omega_{0}(\xi)/\delta\omega_{ 0}(0)$ 
  as a function of the dimensionless variable $\xiti={\xi\Rtfu}/{\aosc^2}
\sqrt{2}$ 
in the case of the surface excitations with $l=1$ and $l=2$. 
The shift decays rapidly with increasing $\xiti$.  
However, as we have anticipated above, the approximation introduced in order to derive 
the hydrodynamic equation~(\ref{eq:hydrodeqtrisxi}), limits the range of validity of
our theory to random potentials with coherence length $\xi$ much smaller than the
healing length $\xi_{\rm heal}=1/\sqrt{8\pi n({\bf 0})a}$ of the superfluid.  
By using the relation between the central density $n({\bf 0})$ and the Thomas-Fermi radius
we have that $\xi_{\rm heal}=\Rtfu(\aosc/\Rtfu)^2$ and
the inequality $\xi\ll\xi_{\rm heal}$ can thus be rewritten as $\sqrt{2}\xiti\ll 1$.
Therefore, the results of Fig. \ref{figure1} for the region $\sqrt{2}\xiti \agt 1$
can only be considered as an extrapolation.

Experimentally, 
the effects of a random potential on the dipole and quadrupole modes of a trapped Bose-Einstein gas
have been investigated in Ref. \cite{Lye05}
by using optical laser speckles 
as realizations of disorder configurations. 
In these experiments, 
the smallest length scale of the speckle potential is of about $\xi\simeq 10 ~\mu {\rm m}$
while the axial Thomas-Fermi radius is $100 ~\mu {\rm m}$.
As a typical experimental situation where the Thomas-Fermi condition 
$N a/\aosc\gg 1$
is satisfied, we can assume $\Rtfu/\aosc\simeq10$.
This would imply a value $\tilde{\xi}\simeq 6$ in our description, which is
far beyond its range of validity.
In particular, for large values of the disorder correlation length, such that $\xi\agt\xi_{\rm heal}$, 
the system 
is not self-averaging and 
the observed frequency-shifts 
must depend on each individual realization of the speckle potential 
\cite{Modugno06,Kuhn07}.
In that case, the disorder average 
must experimentally be taken  by determining, for each mode, the mean value of the
measured  frequencies over 
many different realizations.
Both experiment~\cite{Lye05} and theory~\cite{Modugno06}  predict no shift in that regime
in qualitative agreement with our extrapolation.
Decreasing further the correlation length of the disorder would allow
to enter the regime $\xi\ll\xi_{\rm heal}$, where the system becomes self-averaging,
and the shift predicted in our theory should become observable.

Differently than for the surface modes, the mean-field frequency 
$\omega_{\rm M}=\sqrt{5}\wo$
of the lowest compressional mode
($n_r=1$, $l=0$) is shifted by the beyond mean-field corrections in the normal interaction.
The shift has been calculated in Ref. \cite{Pitaevskii98} and is 
$\delta\omega_{\rm M}/\omega_{\rm M}=\left(63\pi/128\right)\sqrt{a^3 \nzz}$.
Inserting the density oscillation 
$\deltan  \propto1-\left(5/3\right)r^2$ the analogous correction due to the disorder
calculated from Eq.~(\ref{eq:finaleq})
in the limit $\xi=0$ amounts to
${\delta\omega_{\rm M}}/{\omega_{\rm M}}=-({469\pi}/{768})\Rkz'\left(\bf{0}\right)$.

\section{Anisotropic trap}  
\begin{figure}
\begin{center}
\vskip 1 cm
\resizebox{7cm}{4.6cm}{
 \includegraphics{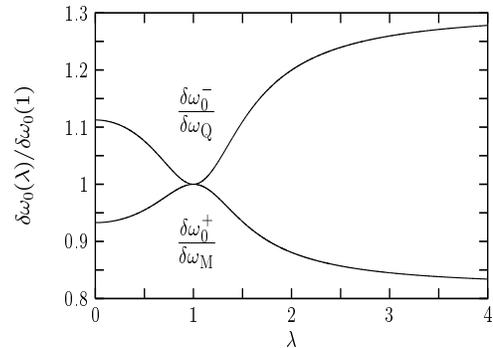}
 } 
\end{center}
  \caption{Relative frequency shift  $\delta\omega_{0}^{+}(\lambda)/\delta\omega_{\rm M}$ (lower curve)
  and
  $\delta\omega_{0}^{-}(\lambda)/\delta\omega_{\rm Q}$ (upper curve)
  of the $m=0$ modes, resulting from the coupling between the quadrupole and monopole modes,
  as a function of the deformation parameter $\lambda=\omega_z/\omegaperp$.  
  The disorder correlation length is taken here $\xi=0$.
  }
  \label{figure2}
\end{figure}

The results can be generalized to non-spherical traps,
by considering 
an axially deformed trap of the form
$\Vext=m\left(\omegaperp^2 \rperp^2+\omega_z^2 z^2
\right)/2$, where $\rperp=\sqrt{x^2+y^2}$ is the radial coordinate. 
In order to put better in evidence the role of the anisotropy, 
we restrict ourselves to the case of disorder with vanishing correlation length.
In the presence of anisotropy, the functions of the form $\deltan\sim r^l Y_{lm}$
are still solutions of the l.h.s. of Eq.~(\ref{eq:hydrodeqtrisxi}).
In the case of the dipole oscillation ($l=1$, $m=0$) with the mean-field mode frequency
$\omega_{\rm dip}=\omega_z$, the disorder corrections lead to the same relative 
shift $\delta\omega_{\rm dip}/\omega_{\rm dip}$
as in the case of the isotropic trap. The same argument applies to the quadrupole mode
$l=2$, $m=2$, with $\omega_{\rm Q}=\sqrt{2}\omegaperp$. 
The quadrupole mode with $m=0$ involves a mixing with the monopole mode
$n_r=1$, $l=0$. In this case the dispersion law is \cite{Stringari96}
$[{\omega_{0}^{\pm}}(m=0)]^2=\omegaperp^2(2+\frac{3}{2}\lambda^2\pm
\frac{1}{2}\sqrt{9\lambda^4-16\lambda^2+16})$, where $\lambda=\omega_z/\omegaperp$
characterizes the deformation of the trap. Using the corresponding oscillation 
$\deltan\sim-\frac{2\mTFu}{m\omegaperp^2}
[(\frac{{\omega_{0}^{\pm}}}{\omegaperp})^2-2]
+\rperp^2+[(\frac{{\omega_{0}^{\pm}}}{\omegaperp})^2-4]z^2$ in Eq.~(\ref{eq:finaleq}) 
we find the shifts
\begin{align}
\label{eq:anisshift}
\frac{\delta\omega_{0}^{\pm}}{\omega_{0}^{\pm}}=-7\pi\Rkz'({\bf 0}) 
\frac{\pm72\pm9\lambda^2+107\sqrt{16-
16\lambda^2+9\lambda^4}}{1536\sqrt{16-16\lambda^2+9\lambda^4}}.
\end{align}
Note that for a spherical trap ($\lambda=1$) we have $\omega_{0}^{+}=\sqrt{5}\wo$
and $\omega_{0}^{-}=\sqrt{2}\wo$ and Eq.~(\ref{eq:anisshift}) recovers the shift
of the quadrupole mode 
${\delta\omega_{\rm Q}}/{\omega_{\rm Q}}=-({35\pi}/{96})\Rkz'\left(\bf{0}\right)$
and the shift 
of the monopole mode
${\delta\omega_{\rm M}}/{\omega_{\rm M}}=-({469\pi}/{768})\Rkz'\left(\bf{0}\right)$
calculated above.
The $\lambda-$dependence of the frequency shifts 
of Eq.~(\ref{eq:anisshift})
is shown in Fig. \ref{figure2}
by plotting the functions 
$\delta\omega_{0}^{+}(\lambda)/\delta\omega_{\rm M}$ (lower curve) and
$\delta\omega_{0}^{-}(\lambda)/\delta\omega_{\rm Q}$ (upper curve).

\section{Conclusions and outlook}  
We have calculated the shifts to the collective frequencies
of a zero-temperature trapped Bose gas induced by a weak external random potential
with a correlation length smaller than the healingh length of the superfluid.
We have shown that the realization of such a limit in current experiments in trapped Bose-Einstein
condensates in random media would allow to measure  
for the first time the 
predictions of the Huang and Meng theory in a quantitative way.
Moreover, the interplay between the pinned and the unpinned components of the condensate 
is expected to  
produce a deviation from the generalized Kohn's theorem
for the center of mass motion of the superfluid. This latter phenomenon 
provides an unambiguous signature of the disorder-induced effects against the beyond mean-field corrections
due to the interatomic interactions.

These results have been derived by means of a hydrodynamic approach and could be extended to consider
the strongly anisotropic traps which realize the so-called
``one-dimensional mean-field" regime described in Ref. \cite{Menotti02}.
This would be relevant in connection with the recent experimental and theoretical
investigations of the transport of Bose-Einstein condensates in one-dimensional  
microtraps in the presence of disorder
\cite{Sanchez-Palencia05,Fort05,Paul04,Paul05,Wang06}.

We thank P. Navez, J. Anglin, H. T. C. Stoof, P. Schlagheck and S. Stringari for useful discussions.
This work was supported by the German
DFG Research Program SFB/TR 12.  

\bibliographystyle{apsrev}

\end{document}